\documentstyle[11pt,aaspp]{article}

\tighten
\received{ 22 April 1996}
\slugcomment{To appear in {\it The Astrophysical Journal, v. 474}}

\newcommand{\rf}{\reference}
\newcommand {\cm}{${\rm cm^{-2}}$}
\newcommand {\kms}{${\rm km\ s^{-1}}$}
\newcommand {\lm}{$\lambda$}
\newcommand {\etal}{et~al.}
\newcommand {\lya}{Ly$\alpha$}
\newcommand {\lyb}{Ly$\beta$}
\newcommand{\civ}{C~{\sc iv}}
\newcommand{\ovi}{O~{\sc vi}}
\newcommand{\he}{He~{\sc ii}}
\newcommand{\neviii}{Ne~{\sc viii}}

\newcommand{\nv}{N~{\sc v}}

\def\lessapp{{~^{\displaystyle <}_{\displaystyle \sim}~}}
\begin{document}

\title{A Composite HST Spectrum of Quasars\footnotemark[1]}

\author{Wei Zheng, Gerard A. Kriss, Randal C. Telfer, John P. Grimes\\
and Arthur F. Davidsen}
\affil{Center for Astrophysical Sciences, The Johns Hopkins University\\
Baltimore, MD 21218--2695\\E\_mail address:
zheng,gak,telfer,grimes,afd@pha.jhu.edu}

\footnotetext[1]{Based on observations with the NASA/ESA Hubble Space
Telescope, obtained at the Space Telescope Science Institute, which is
operated by the Association of Universities for Research in Astronomy,
Inc., under NASA contract NAS5--26555.}

\begin{abstract}

We construct a composite quasar spectrum from 284
HST FOS spectra of 101 
quasars with redshifts $ \rm z > 0.33$. The spectrum covers the wavelengths
between 350~\AA\ and 3000~\AA\ in the rest frame, with a peak S/N level
of $\sim 130$ per \AA\ at $\sim 1200$~\AA. Since $\sim 90\%$ of the sample
quasars have redshift $\rm z < 1.5$, the spectrum is suitable for
studying the wavelength region shortward of \lya\ without
large effects from intervening \lya\ forest absorption.
Data in the waveband between 350~\AA\ and 600~\AA\ are mainly from the
spectra of $\rm z > 1.5$ quasars, for which significant corrections
for the accumulated Lyman--series line and continuum absorption have been
applied.

There is a significant steepening of the continuum slope around 1050~\AA.
The continuum between 1050~\AA\ and 2200~\AA\ can be modeled as a power
law $f_\nu \propto \nu^\alpha$ with $\alpha = -0.99 \pm 0.05$. For the full
sample the power--law index in the extreme ultraviolet (EUV) between 350~\AA\
and 1050~\AA\ is $\alpha = -1.96 \pm 0.15$. For the radio--loud subsample
(60 objects), the EUV spectral index is $\alpha \simeq -2.2$, while for the
radio--quiet subsample (41 objects) it is $\alpha \simeq -1.8$.
The continuum flux in the
wavelengths near the Lyman limit shows a depression of
$\sim 10\%$.  The break in the power--law index and the slight depression of
the continuum near the Lyman limit are features expected in Comptonized
accretion--disk spectra. Comptonization produces a power--law tail in the
wavelength band shortward of $\sim 1000$~\AA\ and smears
out the Lyman--limit edge of the intrinsic accretion--disk spectrum.

In the EUV waveband, we detect several possible emission features, including
one around 690~\AA\ that may
be O~{\sc iii}+N~{\sc iii} produced by the Bowen fluorescence effect.
Comparing our composite spectrum with one made at higher redshifts by
Francis \etal\ (1991), we find that
the equivalent widths of \lya\ and high--ionization emission lines are larger
in our sample, reflecting a known luminosity dependence. The equivalent
widths of low--ionization lines do not exhibit such a dependence, suggesting
that the quasar EUV continuum above $\sim 50$~eV is steeper at higher
luminosity.
Radio--quiet quasars appear to show a slightly harder continuum and lower
ionization levels in their emission lines.

\end{abstract}


\section{INTRODUCTION}

Three decades after their first identification (Schmidt 1963)
more than 7000 quasars have been found (Hewitt \& Burbidge 1993).
Quasar spectra play a key role in observational cosmology.
Most of them have been obtained with ground--based
telescopes using various techniques, including wide--field prism surveys
and spectroscopy at very high resolution.
Despite differences in detail between individual objects,
quasar spectra bear a strong family resemblance, including features such as
a non--thermal power--law continuum, a UV excess, and  broad emission lines.
One way to study the general properties of quasars is by co--adding the
spectra of a number of quasars. The first composite spectrum included
only 14 quasars (Baldwin 1977$a$), while the latest ones (Boyle 1990;
Francis et~al. 1991)  were made from several hundred optical spectra.
High S/N composite spectra serve as a good reference for the
intrinsic continuum shape and emission lines of quasars in the
wavelength region longward of 1200~\AA\ (in the quasar rest frame, and the
same hereafter unless stated otherwise). Ground--based telescopes cannot,
however, reveal the far--UV spectral region below 1200~\AA\ for quasars
with $\rm z < 2$.

Several space--borne telescopes have filled this observational gap and further
extended the observations of quasars to the extreme--ultraviolet (EUV)
region. As of 1995 December, the International Ultraviolet Explorer (IUE)
database contains more than 1500
low--resolution ($R \approx 300$) spectra of 158 quasars.
An extensive IUE--based study of the quasar continuum was carried out by
O'Brien, Gondhalekar, \& Wilson (1988, OGW hereafter).
Using $\sim 200$ IUE spectra of 59 quasars plus some data from
ground--based observations, they found a considerable and gradual
steepening of the continuum toward shorter wavelengths. The power--law index
increases by $\sim 1$ from $1\ \mu \rm m$ to $\sim 800$~\AA. They derived an
average power--law index ($F_\nu \propto \nu^\alpha$) of
$\alpha = -0.67$ between 1216~\AA\ and 1900~\AA\ and $\alpha = -2.36$ between
800~\AA\ and 1216~\AA.

The Hubble Space Telescope (HST) 
has greatly extended our probe of the distant universe. The HST Faint
Object Spectrograph (FOS) database is superior
to the IUE database in terms of S/N level and spectral resolution
(typically $R \approx 1500$). More than 800 FOS spectra of
246 
quasars have been archived as of the end of 1995. The average redshift of
quasars in the FOS database is significantly higher than in the IUE database.
The spectral properties
of some low--redshift quasars have been reported as part of the HST Key Project
``Quasar Absorption Lines'' (Bahcall \etal\ 1993;
Laor \etal\ 1994, 1995). We therefore focused on the FOS archival data for
quasars of higher redshift, in an attempt
to study the sub--\lya\ region in quasar spectra.
We have constructed a composite spectrum of quasars, based
on 284 FOS spectra of 101 quasars. It covers a range between 350~\AA\ and
3000~\AA\ and reveals some features which
are not seen in individual spectra. In particular, we find that the far--UV
continuum slope is significantly different above and below $\sim 1000$~\AA.

\section{DATA REDUCTION}

Archival FOS spectra were retrieved from the Space Telescope Science
Institute in the form of calibrated data.
The selection criteria were (1) target redshift $\rm z > 0.33$; and (2) no
obvious broad absorption lines. The low--redshift limit was set so that the
intrinsic spectral
region shortward of 912~\AA\ is accessible with HST. Our sample includes
55 spectra with grating G160L, 29 with G130H, 93 with G190H, 90 with G270H,
and 17 with G400H. The G160L and G130H spectra were checked for proper
background subtraction and reprocessed if necessary. We measured the
average count level at the wavelengths below 1150~\AA\ where the instrumental
sensitivity should be zero, and used it to correct the background subtraction.
The distribution of the sample with redshift is displayed in Fig.~1.
About 90\% of our sample have $\rm z < 1.5$. In this redshift range,
the correction for \lya\ forest--line absorption is small ($\lessapp 0.1$).
Despite their small number, the spectra of $\rm z > 1.5 $ quasars are
important as they reveal the waveband shortward of 600~\AA. In these quasars,
numerous \lya--forest and metal absorption lines significantly alter the
intrinsic spectra.

The selected FOS spectra were corrected for Galactic extinction.
The hydrogen column density toward each quasar was
estimated using the task {\sf colden} of the Einstein On--Line Service,
Smithsonian Astrophysical Observatory, which is based on the work of Stark
\etal\ (1992). For several southern objects, the column densities were
interpolated from the results of Heiles \& Cleary (1979).
The $E_{B-V}$ values were calculated using the formulation of
Burstein \& Heiles (1978). We used the Seaton curve (Seaton 1979)
to correct for extinction.
The spectra were then shifted to the objects' rest frame.
Data points in the windows near the significant airglow lines at 1216~\AA\
and 1304~\AA\ in the observer's frame were replaced by values linearly
interpolated from neighboring
parts of the spectrum. If a spectrum contained an obvious flux
discontinuity due to absorption by an external Lyman--limit system,
only the undepressed portion of the spectrum was used. Every spectrum has
been visually checked for cosmic--ray spikes, damped
\lya\ absorption lines, or densely clustered absorption lines.
The data points affected by these abnormal features were not used in the
combining process.

In the quasar rest frame, each spectrum was resampled with 0.1~\AA\ bins,
roughly corresponding to the rest--frame bin size
of the original G130H pixels in the highest redshift quasars.
If a new bin is entirely inside an old bin, the flux of the
new bin is taken from the old one. However, if the new bin overlaps two of
the old bins, then the flux is taken as the weighted mean of the fluxes
from the two old bins. The error was treated in the same way, along with two
additional factors: (1) the weighted mean was calculated using the weighting
$1 / e^2$, where $e$ is the error, and (2) the error was multiplied by
$\sqrt {p / 0.1 (1+z) }$ where $p$ is the original data pixel in units
of \AA. While this propagates the magnitude of errors correctly, the
rebinning process does introduce correlated errors between adjacent pixels
that we do not track.

The EUV continuum of high--redshift quasars is unavoidably depressed by the
Lyman--line absorption and the Lyman continuum absorption of
intervening clouds. The accumulated absorption produces a trough
centered around 700~\AA, commonly referred to as
the Lyman valley (M{\o}ller \& Jakobsen 1990). The exact shape of the
Lyman valley depends on the number of absorption lines and their distribution.
We  generated correction curves  using the empirical
formula for the forest--line distribution, i.e.
\begin{equation}
{{\partial^2n} \over{ \partial z \partial N} }
= A (1 + z)^\gamma N^{-\beta}
\end{equation}
(Press \& Rybicki 1993).  For $\rm z > 2$ we chose
$\beta = 1.5$, $A=2.3 \times 10^7$ and $\gamma = 2.46$ because,
with these parameter values, we were able to produce a line number comparable
to that in the recently published data for high--redshift quasars (Hu \etal\
1995). At lower redshifts ($\rm z \approx 0.5$), Kulkarni \& Fall (1993)
found $\beta = 1.48$, $A = 5.5 \times 10^7$ and $\gamma = 0.21$. Since
these two sets of parameters yield a negligible difference in the average
opacity at low redshifts, we used only the parameters derived at $\rm z > 2$.

Using the formulation for the average line--absorption opacity
\begin{equation}
 \tau ( z ) = {{dn} \over {dz}} {{< W_\lambda >} \over {\lambda}} (1 + z),
\end{equation}
where $dn / dz$ is the number of lines per unit redshift and $< W_\lambda >$
is the mean rest--frame equivalent width of \lya\ lines averaged over
the assumed distribution in column density (M{\o}ller \& Jakobsen 1990),
we calculated the wavelength--averaged opacity produced by Lyman-series
lines up to level 50. The accumulated Lyman--continuum absorption was
calculated using $\tau (\lambda)= N  \sigma_0 \ (\lambda/\lambda_0)^3$,
where $N$ is the neutral hydrogen column density and $\sigma_0$ is the
hydrogen
photoionization cross section at the Lyman--limit wavelength $\lambda_0$.
We produced a set of correction curves at different redshifts
for lines with column density between $2 \times
10^{12}$ and $10^{16}$~\cm.
The low limit was set by the recent Keck spectroscopic observations (Hu \etal\
1995). Absorption by any cloud with column density greater than $10^{16}$~\cm\
produces a noticeable Lyman--limit edge in a spectrum, and
its correction is straightforward.
Fig.~2 displays correction curves at different redshifts and clearly shows
that the correction is small ($\lessapp 0.1$) for $\rm z < 1.5$.
We first applied a general correction for every merging spectrum.
For quasars with $\rm z > 1.5$, if the absorbers with
column density $N > 10^{16}$~\cm\ are known in the
literature, we applied  additional corrections for them.

To merge the quasar spectra we used a technique similar to that described by
Francis \etal\ (1991). We sorted the quasar spectra by their wavelength
coverages. Starting from the
quasar spectrum that covers the longest wavelengths,
we normalized the flux of the second
spectrum to the first so that the average flux of the two spectra matched
each other in the overlapping region.  To assure that the normalization
applied to the continuum only, wavelength windows surrounding major emission
lines were excluded. Using this new summed spectrum, the next
spectrum was normalized and merged. This process was repeated to include
more spectra, until the spectrum that covers the shortest wavelengths
was included. We also practiced this method starting at the
shortest wavelength end and found that the overall continuum and emission--line
features were virtually identical. However, the summation from longer
wavelengths is preferred, as it follows the direction from high to low
S/N level at shorter wavelengths.

Throughout the combining process the weighted mean and
the error in the weighted mean were calculated at each 0.1~\AA\ pixel.
We tried a weight $w = f/e^2$, where $f$ is the flux and $e$ the error.
With such a weighting a flux spectrum would be converted into an approximate
count spectrum. Such a summation  resembled the addition of raw count
spectra and produced an output with the highest possible S/N level (with a
peak value of $\sim 300$ per \AA). However,
a small number of high--quality spectra were given particularly high weights
so that the final spectrum bears the signature of strong absorption lines in
these few quasars.
We have, therefore, chosen a uniform weight in order to produce a
smoother spectrum, particularly at shorter wavelengths, more representative
of the overall quasar population.
For every data point of every merged spectrum, the
average S/N level was calculated over 10 points (1.0~\AA) on either side.
If the average S/N level per pixel was below 0.3 this data point was excluded.

Using this composite spectrum as a template, we further applied a
renormalization process. All the spectra were normalized so that the average
flux level in individual spectra, excluding the wavelength windows surrounding
major emission lines and the data points with average S/N
level less than 0.3, matched that of the template spectrum in the
corresponding wavelength range. The sum of the normalized spectra was smoother
than the template at shorter wavelengths where the S/N level was lower,
and it serves as the final composite spectrum.

Fig.~3 displays the number of merged spectra as a function of wavelength.
For some objects, multiple observations taken with the same grating
exist. In such cases, we merged these spectra into a single spectrum, which
counts as one in Fig.~3.
Near 1200~\AA\ the number of merged spectra reaches a peak of $\sim 100$,
and at very short wavelengths it is only 2 or 3.
Fig.~4 displays the S/N level of the
composite spectrum after binning by 10 pixels.
The spectral region near 1200~\AA\ has the highest S/N level of $\sim 130$
per \AA, and the region near 350~\AA\ has $\rm S/N \approx 8$ per \AA.
The actual S/N level may be slightly lower as the flat field correction
of the FOS instrument limits the S/N in any given spectrum to $\sim 50$ or
less. But we do not expect a significant effect as any flat--field
irregularities should be smeared out due to the different redshifts.
\newpage
\section{RESULTS}

\subsection{Continuum}

The composite HST quasar spectrum, as shown in Fig.~5,
shows a change in slope of the continuum between 350~\AA\ and 3000~\AA\
that can be approximated as a broken power law.
We used the IRAF task {\sf specfit}
(Kriss 1994) to fit the shape of the continuum and the emission lines.
With {\sf specfit}, complex line and continuum models can be fitted to data
using a nonlinear $\chi^2$-minimization technique.
For our fits we binned the composite spectrum by three pixels
(0.3~\AA).
The following wavelength windows for the continuum fitting were selected:
350--450, 500--640, 720--750, 800--825,
930--950, 1100--1130,
1450--1470,
1975--2010, 2150--2200~\AA.
Above 900~\AA\ any small deviations from a power law that are due
to weak emission features, Fe~{\sc ii} in particular,
yield significant $\Delta \chi^2$ because of the very high S/N level.
Therefore, the selected windows at long wavelengths are narrow.
The region near 912~\AA\ was excluded because of an apparent local depression
in flux (discussed below).
The region longward of 2200~\AA\ was not fitted because of the strong
Fe~{\sc ii} emission seen there.

Since the propagated error arrays in our calculations contain correlated
errors that are not taken into account, we experimented with two different
methods of assigning errors and evaluating the goodness of our fits.
The first directly uses the propagated error array.
The second uses the root-mean-square (rms) spectrum.
An advantage of using the rms spectrum for the error bars is that
it avoids giving excessive weight to the longer-wavelength portions of the
spectrum where the number of merging spectra is large.
To calculate the rms spectrum,
each normalized spectrum was differenced with respect to the composite.
The squared differences were summed, with equal weights, and the sum was
divided by the number of spectra contributing to each pixel.
The square root of the result gives the rms deviation,
shown in Fig.~6. The dispersion is $\sim 10\%$ around 2500~\AA, and
$\sim 30\%$ around 400~\AA.

A fit with a single power law continuum using the propagated error array
yields $\chi^2 = 13,684$ for 1504 data points and 2 free parameters.
In contrast, a broken power law gives $\chi^2 = 1196$ for 1504 points and
4 free parameters, with the power--law index $\alpha = -1.00 \pm 0.01$
above 1038~\AA\ and $\alpha = -2.02 \pm 0.02$ below.
We conclude that a broken power law is
a good fit to the continuum (except for the region between 900~\AA\ and
1200~\AA\ where the continuum shape changes gradually),
and a single power law can be excluded at
much greater than the $5 \sigma$ confidence level.

Fits using the rms spectrum for the errors give similar results.
The fitted power--law index $\alpha = -0.99 \pm 0.01$ above
$\sim 1052$~\AA\ and $\alpha = -1.96 \pm 0.02$ below.
The power--law index in the region with $\lambda >
1052$~\AA\ derived with the rms array is the same as
that with the propagated error array, and it is slightly different
for $\lambda < 1052$~\AA. Therefore
the power--law indices in Table~1 are not affected significantly by the
use of different error arrays, and we feel that using the rms array may yield
more representative results at shorter wavelengths.
Note that the $1 \sigma$ errors
only reflect the formal estimates of the fitting process, and
the actual errors are somewhat higher because of the uncertainties introduced
in the merging process.

The break point of the power law is $1052 \pm 13$~\AA.
This value is sensitive to the fitting windows chosen at shorter wavelengths,
and may vary by $\sim 50$~\AA. The continuum fit parameters are
summarized in Table~1.
Separate fits were also performed for radio--loud and radio--quiet quasars, and
these results are also given in Table~1 and discussed below.
As one can see in Fig.~5, one power law can describe well the wavelengths
between 1300~\AA\ and 3000~\AA, despite the presence of Fe~{\sc ii} emission.

The spectral region near the Lyman limit is of special interest.
We plot this region in Fig.~7 with a pixel size of 0.3~\AA.
With respect to the fitted broken power--law continuum, there is a $\sim 10\%$
trough just below the Lyman limit.
The trough is a smooth feature
centered around 870~\AA, with a wavelength span of $\approx 70$~\AA.
It does not seem to arise from
associated absorbers because it recovers much faster at shorter wavelengths
than a $\tau \propto (\lambda / \lambda_0)^3$ law predicts.
If included in the fit, this trough contributes substantially to $\chi^2$
as shown in the lower panel of Fig. 7.
Our fits to a Comptonized accretion disk spectrum, discussed in \S 4.4,
address the statistical significance of the trough in more detail.
Such a trough is seen in the merged spectra made of several subsamples and
therefore is not an artifact due to a few individual quasars.
There are several spectra that show significant broad absorption
features near the Lyman limit, and they will be discussed in a
separate paper (Zheng \etal\ 1996).

\subsection{Emission Lines}

We used multiple Gaussian components to fit the emission lines using
{\sf specfit}.
\lya, \civ, \ovi, \nv\ and Mg~{\sc ii} were fitted with two Gaussians:
a narrow component and a broad one. The line widths of the two \nv\
components were tied to the respective \lya\ components, and the
\nv/\lya\ wavelength ratio was fixed at its rest--frame value.
Other lines were modeled with one Gaussian. The line widths of
Al~{\sc iii}, Si~{\sc iii}] and C~{\sc iii}] were tied together in the
fitting process.
The fitting results are listed in Table~2. The flux units are arbitrary on a
relative scale with I(\lya) = 100. In addition to the major emission
features marked in Fig.~5, some weak features, such as C~{\sc iii} \lm 977
and N~{\sc iii} \lm 991, can be identified because of the high S/N level.
The emission feature around 780~\AA\ has been identified
as Ne~{\sc viii} (Hamann, Zuo, \& Tytler 1995). The ionization
level of Ne~{\sc viii} (206~eV) is extremely high. If the line emission
is generated by collisional excitation, a
high temperature ($\rm T > 5 \times 10^4$~K) is needed in the region in which
these lines are formed as their excitation energies are greater than 12~eV.
The \neviii\ profile, with a FWHM $\approx 12000$~\kms, is much broader
than that of most other
lines, and \ovi\ emission also exhibits a broad base of similar width.
The emission feature around 690~\AA\ may be O~{\sc iii}+N~{\sc iii}
arising from the Bowen fluorescence effect (Eastman \& MacAlpine
1985; Netzer, Elitzur, \& Ferland 1985). The O$^{++}$ and N$^{++}$ ions
have transitions whose wavelengths coincide very closely with \he\ \lm
304~\AA, which was recently detected in a high--z quasar
(Davidsen, Kriss, \& Zheng 1996$a$). Abundant \he\ \lya\ photons in the
line--emitting region can excite the O$^{++}$ and N$^{++}$ ions and produce
cascade
transitions. Theoretical estimates  suggest that the O~{\sc iii}/\he\ \lya\
intensity ratio is on the order of 0.1 and is only weakly dependent on the
ionizing flux.  This feature is not present in every spectrum that covers
the wavelengths between 650~\AA\ and 700~\AA, so its identification is only
suggestive.
There is also an emission feature near 833~\AA. We tentatively identify it as
O~{\sc ii} \lm 833.80, while an alternative identification as
O~{\sc iii} \lm 833.97 is also possible. At wavelengths shorter than 600~\AA\
we did not attempt to identify emission features because of the limited S/N
level.

\section{DISCUSSION}

\subsection{EUV Continuum Shape}

The continuum shape above 600~\AA\ can be determined quite accurately,
because the spectrum is composed of mainly $\rm z < 1.5$ quasars
and the Lyman--valley correction is small. Below 600~\AA, the merged
spectra are mainly derived from  $\rm z > 1.5$ quasars, and the results
depend more and more on the correction as the redshift increases.
The correction curves applied are based on Equation (1), which is statistical
in nature. Our Monte--Carlo simulations of forest--line absorption (with
column density $N < 10^{16}$~\cm) suggest
fluctuations of $\sim 10\%$ in the average decrements.
We find that the actual average line opacity, due to line blending, is
$\sim 15\%$ smaller (for $z \sim 3$) than Equation (2)
indicates. As lines with higher column density contribute more
significantly in the wavelengths around 700~\AA, the correction for lines
between $10^{16}$ and $10^{17}$~\cm\
introduces larger errors. Our calculations show that, at $\rm z \sim 2$,
absorption lines with $10^{16} < N < 10^{17}$~\cm\ may affect the decrement
at 350~\AA\ and lead to an uncertainty in the power--law
index between 350~\AA\ and 1050~\AA\ of $\sim 0.15$. However,
such a large error is unlikely because an absorption line
with $N \geq 10^{16}$~\cm\ should produce a visible Lyman--limit edge, and
our selection process has carefully excluded the wavelength regions shortward
of such visible edges. Combining all these facts together, we feel
that an uncertainty of 0.15 in the power--law index may be a reasonable
estimate
between 350~\AA\ and 1050~\AA. At wavelengths above 1050~\AA, a comparison
between different quasar subsamples yields a dispersion in $\alpha$ of
$\sim 0.05$.

Our measurements of the quasar continuum can be compared with earlier IUE
results.
OGW (1988) found the continuum steepening toward shorter wavelengths.
Excluding blazars and broad--absorption--line quasars,
their power--law indices are
$\alpha = -0.67 \pm 0.05$ for $\lambda > 1216$~\AA, $-2.14 \pm 0.13$
for $1216 > \lambda > 912$~\AA,
and $-2.36 \pm 0.21$ for $1216 > \lambda > 800$~\AA.
The far--UV power--law index they derived is somewhat steeper than our
result of $\alpha \approx -2.0$.
Their subsample with $\rm z > 1.1$ yielded even steeper values at
shorter wavelengths, i.e.
$\alpha = -0.46 \pm 0.05$ for $\lambda > 1216$~\AA, $-2.57\pm 0.15$ for
$1216 > \lambda > 912$~\AA, and $-2.86 \pm 0.24$ for $1216 > \lambda >
800$~\AA.
The OGW results are in qualitative agreement with ours,
but our larger sample and more sensitive HST observations
more accurately determine the power--law index.
OGW (1988) suggested that the steepening is gradual
from the optical to the UV.
We find that the continuum between
$\sim 1200$~\AA\ and 3000~\AA\ can be approximated with one power law.

At shorter wavelengths, O'Brien (1987) used the IUE spectra of three
radio--quiet quasars to derive a mean power--law
index of $\alpha = -1.77 \pm 0.46$ between 600~\AA\ and 912~\AA. Our
more precise measurement, while consistent with his result,
suggests that for our total sample the quasar continuum in this spectral
region is actually slightly steeper. For our radio--quiet subsample, however,
we find $\alpha$ very similar to O'Brien's value, but with a considerably
smaller uncertainty.

EUV photons play important roles in many physical processes, and knowledge of
the quasar EUV continuum shape would lead to improvements in many theoretical
models. Emission lines in active galaxies and quasars are believed to be
powered by photoionization. Models with various EUV continuum shapes
yield different line intensities (Krolik \& Kallman 1988).
A typical AGN power--law index adopted for the EUV band is $\alpha \approx
-1.0$
(Mathews \& Ferland 1987) which predicts too strong an ionizing continuum
compared to our result of $\alpha \approx -2.0$.
Our composite spectrum, with a power--law extrapolation into the soft--X--ray
band, can serve as a reasonable input for photoionization models in the
study of AGN.

Quasar radiation plays a key role in the reionization of the
intergalactic medium in the early universe (e.g., Shapiro, Giroux, \& Babul
1994).
The exact quasar continuum shape (Miralda--Escude \& Ostriker 1990) may dictate
the \he--ionizing metagalactic radiation field, which in turn affects the
\he\ opacity in the intergalactic medium. The recent measurement of the
\he\ opacity (Davidsen \etal\ 1996$a$) in the intergalactic medium
is compatible  with a background radiation field with $\alpha \approx -1.8$
(Davidsen \etal\ 1996$b$).
This is very similar to the value found here for radio--quiet quasars, which
should dominate the metagalactic background radiation.
The continuum shape of quasars also affects their detectability at high
redshifts. An order of magnitude fewer quasars are predicted to be
observable in the UV band if their EUV continuum power law index is
$\alpha = -2$ compared with $\alpha = -1$ (Pickard \& Jakobsen 1993).

\subsection{Relation with Radio Properties}

We divided our sample into radio--loud and  radio--quiet objects and applied
the same fitting process to the lines and the continuum. Radio--loud
quasars are defined as those with the logarithm of the ratio of the 6cm flux
to the flux in the optical B band $R_L = \log (f_{6cm}/f_B) > 1$.
To evaluate $R$, we used fluxes for the selected quasars from V\'eron-Cetty
\& V\'eron (1993).
The composite spectra of radio--loud and radio--quiet quasars
are compared in Fig.~8. Both
composite spectra cover a similar wavelength range and exhibit similar
spectral properties. At very short wavelengths the composite spectrum of
radio--loud quasars is entirely due to two objects:
Q0446--2049 between approximately 475~\AA\ and 600~\AA, and OQ~172 below
475~\AA, so the result is quite uncertain at these wavelengths.

For radio--loud quasars, the fitted power--law index $\alpha =
-1.02 \pm 0.01$ above 1050~\AA\ and $-2.16 \pm 0.03$ below.
For radio--quiet quasars, it is $-0.86 \pm 0.01$ and $-1.77 \pm 0.03$,
respectively. Our results are in agreement with Christiani \& Vio
(1990) in terms of the similar continuum shape between
1000~\AA\ and 3000~\AA\ for their subsamples of radio--selected and
optically--selected quasars. Below
1000~\AA, the continuum of radio--loud quasars appears to be steeper
than that of radio--quiet quasars.
Such a difference was also noticed by OGW (1988) in their radio--selected and
optically--or--X--ray selected samples.

For radio--quiet quasars, the Mg~{\sc ii} intensity  
is stronger, and the intensities of \civ\ and \ovi\ are weaker.
Radio--quiet and radio--loud quasars therefore appear to show somewhat
different ionization levels of their
emission lines. To check whether this is
a luminosity effect, we compared the mean redshift of the subsamples.
As shown in
Table~1, these two groups of quasars have similar mean redshifts.
We also calculated the quasar specific luminosity at 1000~\AA, assuming $H_0 =
75 \rm\ km\ s^{-1}\ Mpc^{-1}$ and $q_0 = 0.5$. If a spectrum does not cover
1000~\AA, the flux was extrapolated from that at the wavelengths
closest to 1000~\AA, using the power law derived from the whole sample
(Table~1). The mean luminosities of the radio--loud and radio--quiet quasars
are both $2.4 \times 10^{30}\ \rm erg\ s^{-1}\ Hz^{-1}$.
We therefore conclude that the differences seen are not due to a luminosity
effect.

\subsection{Redshift and Luminosity Effects}

Fig.~9 plots our composite spectrum for radio--quiet quasars and the
spectrum of Francis et~al. in the region of overlap.
At wavelengths longer than 1000~\AA, our spectrum is mainly derived
from quasars with $\rm z < 1.5$, while that of Francis \etal\ is mainly
for $\rm z > 2.5$.
The two spectra are normalized so that the continuum levels around
1450~\AA\ match each other. The high--z spectrum shows a significant
flux depression shortward of \lya\ resulting from \lya\ forest--line
absorption. The average power--law index between 1200~\AA\ and 2200~\AA\ is
$\sim -0.8$,
reflecting a harder UV continuum at higher redshifts. The difference in the
continuum shape does not occur smoothly. The continuum in the two spectra
matches fairly well between 1200~\AA\ and 1900~\AA.
Above 1900~\AA\ the continuum in the high--z spectrum becomes significantly
harder ($\alpha = -0.32 \pm 0.05$). This result supports the finding of OGW
(1988) that quasars of higher redshift show harder UV continua.
Francis (1993) further confirmed that the effect is a correlation with
redshift and not with luminosity.

By comparing the two composite spectra in Fig.~9 one can also see that
the intensities of several high--ionization emission lines are different.
One should approach these differences cautiously, however, since
our composite spectrum is more subject to
selection effects and more heterogeneous than the Francis \etal\ spectrum.
The equivalent widths of \lya, \civ, \he+O~{\sc iii}], and \ovi\
decrease by approximately 39\%, 37\%, 42\%, and 56\%, respectively,
from $\rm z \sim 1$ to $\rm z \sim 3$.
In particular, the \ovi\ intensity is significantly weaker at higher redshifts,
but we note that for \lya\ and \ovi, the Francis \etal\ spectrum is strongly
affected by \lya\ forest--line absorption.
On the other hand, the equivalent widths of low--ionization lines, such as
Mg~{\sc ii}, C~{\sc iii}], O~{\sc i} 
and Si~{\sc iv} 
remain basically unchanged.
The trend that \lya\ and high--ionization lines exhibit significant
decreases toward higher redshifts
(and, consequently, higher luminosities: the Baldwin effect, Baldwin 1977b)
supports the hypothesis that the Baldwin effect is due to a luminosity
dependence of the shape of the ionizing continuum (Zheng, Fang, \& Binette
1992), i.e. the continuum shape above $\sim 50$~eV may be steeper at
higher luminosities.

\subsection{Comptonized Disk Spectrum}

The flux discontinuity at the Lyman limit of 912 \AA\ is a critical test
of the nature of the UV excess in AGN. The rising continuum in the
optical--UV band resembles the thermal spectrum of an accretion disk (Shields
1978; Malkan \& Sargent 1982). Such a disk, with a characteristic temperature
of $\sim 10^4$~K, should produce a significant
absorption edge at the Lyman limit (Kolykhalov \& Sunyaev 1984).
Previous searches for such a feature in the optical and IUE database have
yielded negative results (Antonucci, Kinney, \& Ford 1989;
Koratkar, Kinney, \& Bohlin  1992; Koratkar et~al. 1995).
In general these observations limit any discontinuity due to a Lyman--edge
feature to less than 15\%.

Previous investigations of Lyman--limit discontinuities in quasar spectra
based on HST data have led to different, tentative reports.
Sun, Malkan, \& Chang (1993), analyzing the FOS spectra of 21 quasars,
suggested that $\sim 24$\% of the sample have possible rest--frame
Lyman--limit
absorption edges.  They found similar results for a sample of medium-- to
high--redshift quasars using optical spectra obtained at Lick Observatory
and at the Hale Observatory.
On the other hand, Tytler \& Davis (1993) found no discontinuity across the
Lyman limit in the FOS spectra of 27 quasars.
While it is not clear whether these two conflicting reports are based on
different samples, the larger size of the sample comprising our mean
spectrum assures higher quality statistics for examining the Lyman--limit
region.

Two features in the mean spectrum we have constructed suggest association
with an intrinsic spectral feature in the Lyman--limit region.
The first is the sharp break in the continuum spectral index.
The second is the slight depression in the continuum flux shortward of the
Lyman limit noted in \S3.1.
The continuum break resembles that observed shortward of 1000~\AA\
in the Hopkins Ultraviolet Telescope spectrum of the quasar 3C~273
(Davidsen et~al. 1996$c$).
The change in the spectral index is so abrupt that it cannot be produced either
by reddening or by intervening absorption, including the Lyman--series lines
and
the Lyman continuum (which we have largely corrected for).
Comptonization of an accretion--disk spectrum with an intrinsic Lyman edge
can produce a continuum
shape very similar to what is shown by our mean spectrum in Fig.~5.
It produces a power--law high--energy
tail and smears out any intrinsic Lyman--limit discontinuity
(Czerny \& Zbyszewska 1991).
The emergent spectrum still bears the signature of a Lyman edge, however,
if the optical depth to scattering is not too high.

To fit the overall continuum shape of our mean quasar spectrum, we computed
an intrinsic disk spectrum in the Schwarzschild metric that was a sum of
blackbodies with an empirical Lyman limit feature as described by
Lee, Kriss, \& Davidsen (1992).  A spherical corona with optical depth
$\tau_e = 1.0$ to Compton scattering was assumed to surround the disk.
The Comptonized disk spectrum was calculated using the formulation of
Czerny \& Zbyszewska (1991) as described by Lee \etal\ (1992).
The shape of the intrinsic disk
spectrum is invariant for a fixed ratio $\dot{m} / M_{BH}^2$,
and the spectral index of the Comptonized high energy tail depends almost
entirely on the Compton $y$ parameter $y \propto \tau_e^2 T_c$  (Lee 1995).
The mass accretion rate determines the total luminosity, and this we normalized
to the mean luminosity of the quasar sample.
The location of the Lyman-edge depends largely on the disk inclination, $i$.
Thus the free parameters in the fit were
the mass of the central black hole, $M_{BH}$,
the mass accretion rate, $\dot{m}$, the temperature of the Comptonizing
medium, $T_c$, the optical depth of the empirical Lyman edge, $\tau_{Ly}$,
and the inclination of the disk normal to the line of sight, $i$.

Our fits used the same continuum windows listed in \S 3.1 for the broken
power-law continuum fits except that we also included the Lyman edge region
from
825--930 \AA\ that was excluded in those previous fits.
For error bars we used the propagated errors computed in the merging process.
We find that the best fit requires a Lyman edge of modest optical depth in the
emergent disk spectrum.
The best fit has a Lyman limit optical depth of $\tau_{Ly} = 0.8$ and
yields $\chi^2 = 1715$ for 1853 points and 5 free parameters.
$\tau_{Ly} = 0.0$ results in a fit with $\chi^2 = 1761$, and
this can be excluded at $> 5\sigma$ confidence relative to the fit with a
Lyman edge of finite optical depth.

Fig.~10 shows the best-fit Comptonized accretion-disk spectrum compared to the
1-\AA\ resolution mean quasar spectrum.
This model has $M_{BH} = 1.4 \times 10^9\ M_\odot$,
$\dot{m} = 2.8\ M_\odot~\rm yr^{-1}$, $\tau_{Ly} = 0.8$,
$i = 30^\circ$, and $T_c = 4.1 \times 10^8$~K.
For a Schwarzschild black hole the total luminosity of the disk is
$8.5 \times 10^{45}~\rm erg~s^{-1}$, which is 5\% of the Eddington limit
for a $1.4 \times 10^9\ M_\odot$ black hole.

Although an inclination closer to $60^\circ$ would have been expected
for a sample of randomly oriented disks, we note that the heterogeneous
HST archive sample is unlikely to be random.
Our exclusion of BAL QSOs is likely to avoid disks at high inclinations.
The large fraction of radio-loud objects in the sample will also
produce a bias towards low inclinations.

The geometrically thin, optically thick ``standard" accretion disk model
has been criticized on many counts (e.g., Antonucci 1988, 1992).
Among its shortcomings are the lack of strong Lyman--edge features in
the observations, the difficulty of constructing disks with hot--enough
tails to match the observed soft--X--ray excesses of AGN, and observed
polarization properties that are inconsistent with simple predictions.
Comptonization of the intrinsic disk spectrum in a quasi--spherical
scatterer resolves many of these problems.

As noted by Czerny \& Zbyszewska (1991), Comptonization can broaden an
intrinsic Lyman edge and decrease its contrast to the point where a
broad range of intrinsic disk parameters produce similar emergent spectra.
The ability of such a model to match the weak Lyman limit feature in our
mean spectrum is illustrated by our fit in Fig.~10.

O'Brien (1987) showed that an extrapolation of a $\nu^{-2}$ power law
may connect the UV intensity to the soft--X--ray flux level around 0.3 keV
in many quasars. This implies that, if the EUV continuum in our composite
spectrum is representative, a single power law representing the high energy
tail of a Comptonized accretion--disk spectrum may bridge the gap between
the far--UV and the soft--X--ray bands.
A Comptonized accretion disk fit to the Hopkins Ultraviolet Telescope (HUT)
spectrum of Mrk~335
(Zheng et~al. 1995) matches both the far--UV spectral shape in the
Lyman--limit region and the soft--X--ray excess observed with the
Broad Band X--ray Telescope (Turner et~al. 1993).

If the Comptonizing medium has a largely spherical distribution, this can
resolve many of the problems that recent polarization observations present
for the thin disk model.
Simple disk models show a low polarization increasing to the blue with a
large drop beyond the Lyman limit due to the large absorption opacity of
the Lyman continuum (Laor, Netzer, \& Piran 1990).
This predicted behavior is not observed in recent observations
(Impey et~al. 1995; Koratkar et~al. 1995; Antonucci et~al. 1996) which
generally show little or no polarization and occasionally a {\it rise}
beyond the Lyman limit rather than a drop.  Moreover, the polarization vectors
tend to be parallel to the radio--jet axis rather than perpendicular as one
would expect if the jet were oriented perpendicular to the disk.
A more physically realistic treatment of the polarized radiative transfer
in a thin disk atmosphere (Blaes \& Agol 1996) demonstrates that polarization
can increase across the Lyman limit as observed.
If most of the intrinsic disk radiation is Compton scattered, however,
the polarization properties will be determined by the Compton scattering
medium, not the disk atmosphere.  If the scatterer is spherical, the
polarization will be low.  Antonucci et~al. (1996) note that
the polarization vector will lie parallel to the radio jet
if the scatterer is slightly extended along the radio axis.
This is perhaps not unexpected if a wind from the disk feeds the scattering
region.

Although strong Comptonization can obscure many of the intrinsic properties
of the disk spectrum, it does not destroy them.
The shape of the spectrum near the peak in $\nu f_\nu$ and to the red is still
determined predominantly by the black hole mass and the accretion rate.
The position of any surviving Lyman--edge feature depends on the disk
inclination.
The spectral index of the high energy tail is diagnostic of the optical
depth and temperature in the Comptonizing medium.
Given the strong suggestion that Comptonization dominates the appearance
of our composite HST spectrum, future disk models should take this
feature explicitly into account.

\section{SUMMARY}

We have constructed a composite spectrum from 284 HST/FOS spectra of
101 quasars with $\rm z > 0.33$.
The UV continuum in the composite quasar spectrum can be fitted with a broken
power law with a break around 1050~\AA. At longer wavelengths the
power--law index is $\alpha \simeq -1.0$.
The EUV continuum can be fitted with a power law of $\alpha \simeq -2.0$
without a noticeable turnover toward higher frequencies.
Near the Lyman limit the continuum flux is depressed by $\sim 10\%$ relative
to surrounding wavelengths.
These features match the characteristics of thermal emission from
an accretion disk with Lyman--edge absorption that is modified by electron
scattering, suggesting that
the UV bump may be produced by a Comptonized accretion--disk spectrum.

The strengths of high--ionization emission lines show a significant luminosity
dependence, probably due to evolution of the UV continuum shape. In
contrast, the equivalent widths of low--ionization lines do not exhibit
such a dependence, suggesting that the continuum evolution may take place
mainly in the energy band above 50 eV.
The EUV continuum appears to be steeper in radio--loud quasars ($\alpha \simeq
-2.2$) than it is in radio--quiet quasars ($\alpha \simeq -1.7$).
Radio--quiet quasars appear to show slightly lower ionization levels in
their emission lines.

The accumulated quasar radiation, with a continuum spectrum as found here,
is capable of
ionizing helium in the intergalactic medium and producing an opacity
comparable to that recently
measured with HUT (Davidsen \etal\ 1996$a$). Our findings support models that
attribute metagalactic radiation in the early universe to quasar radiation.

\acknowledgments

Support for this work has been provided by
contract NAS5--27000 and grant NAG5--1630 from NASA, and
grant AR--4389.01--92A and AR--5284.01--93A from the
Space Telescope Science Institute, which is operated by the Association of
Universities for Research in Astronomy, Inc., under NASA contract NAS5--26555.
We thank P. Hewett for providing the ground--based composite spectrum of
Francis \etal\ (1991) in digital form.

\clearpage

\vfill
\begin{center}
{\doublespace
{\large Table 1. \,\, Fitted Power--Law Index$^a$}
\bigskip

\begin{tabular}{|c|ccc|} \hline\hline
Sample   & All Quasars & Radio Loud & Radio Quiet \\ \hline
Number of Objects & 101 & 60 & 41 \\
Mean Redshift & 0.93 &0.87 &0.95  \\
Wavelength Range (\AA) & & & \\ \hline
1050--2200  &$ -0.99 \pm 0.01$ &$-1.02 \pm 0.01$& $-0.86 \pm 0.01$ \\
\phantom{0}600--1050  &$ -2.02 \pm 0.05 $ &$-2.45 \pm 0.05$& $-1.83 \pm 0.03$
\\
\phantom{0}350--1050   &$ -1.96 \pm 0.02$ &$-2.16 \pm 0.03$ &$-1.77 \pm 0.03$
\\
\hline
\end{tabular}
}
\end{center}

\noindent $^a$\,\, The $1 \sigma$ errors are statistical only, obtained from
the error matrix of the fit. Due to
the smaller sample size and the Lyman-valley correction, the
actual errors below 1050~\AA\ are probably on the order of $\pm 0.15$.

\vfill

\clearpage
\begin{center}
{\large Table 2. \,\, Emission Line Strengths in Composite Quasar Spectrum}

{\singlespace
\begin{tabular}{lccccccc} \tableline\tableline
 ~~~~Line & $\lambda_0$ & \multicolumn{2}{c}{Whole Sample}&
\multicolumn{2}{c}{Radio Loud}&\multicolumn{2}{c}{Radio Quiet}\\
 & (\AA) & Flux  & EW (\AA)& Flux &EW (\AA) & Flux &EW (\AA)\\ \hline
  O {\sc iii} + N {\sc iii}$^a$& $\sim 690$&$4.6\pm 1.6$&3.6&5.2&4 &5.1&3.5 \\
  Ne {\sc viii}& 773.71 &$4.0\pm 1.5$ & 3.2 & 3.6 & 3 &4.2& 4\\
  O {\sc ii}$^a$ &833.80 &$1.4\pm 0.3$&1.2 & 0.5& 0.4 & 2.2 & 1.6\\
  C {\sc iii} & 977.02& $0.9\pm 0.3 $ & 0.7 &0.8 & 0.6&1.4 & 1.1\\
  N {\sc iii} &990.98 & $ 1.1\pm 0.4 $&0.8 &0.8 & 0.6 & 1.2& 1.1\\
  \lyb + \ovi &$\sim 1031$&$19 \pm 2$&16& 19 & 17 & 16 &12\\
  Ar {\sc i}$^a$ & 1066.66& $0.6 \pm 0.2$& 0.5 & 0.5 &0.4 &1.3 & 1.0\\
  \lya & 1215.67& $100 $&92 & 100 & 96 & 100& 85\\ 
  \nv & 1240.15 & $11 \pm 1$ & 10 & 10 & 10& 14 & 10\\
  O~{\sc i} & 1303.94 & $1.3 \pm 0.2$ & 1.2 &1.3 & 1.2 &1.1& 1.0\\
  C {\sc ii} & 1334.53 & $0.3\pm 0.1 $  &0.3& 0.2 &0.2& 0.4 & 0.4\\
  Si~{\sc iv}+ O~{\sc iv}] &$\sim 1400$&$7.5\pm 0.5$&8&6.8 & 8 & 8.1 & 8\\
  \civ & 1549.01       & $62 \pm 0.4$& 69 & 66 & 77 & 52 & 59\\
  \he\          & 1640.46 & $3.9\pm 0.2$& 4.5 &4.3& 5 & 3.4 & 3.9\\
  O~{\sc iii}]  & 1664.15 & $2.9\pm 0.3$& 3.4 &3.2 & 3.7 & 2.7 & 3.1 \\
  N~{\sc iii}]  & 1750.46 & $0.5\pm 0.1$& 0.7 & 0.7 & 0.9 & 0.2 & 0.2\\
  Al~{\sc iii}  & 1857.40 & $2.4\pm 0.3$& 3.5 & 2.3 & 3.6 & 2.9& 3.5 \\
  Si~{\sc iii}] & 1892.03 & $2.4\pm 0.3$& 3.5 & 2.3 & 3.5 & 2.8& 3.5 \\
  C~{\sc iii}] & 1908.73& $11.5 \pm 0.3$& 17  & 11 & 17  & 12 & 17\\
  C~{\sc ii}] & 2324.27& $2.9\pm 0.9$ & 5 & 2.9 & 6 & 2.3 & 4 \\
  Fe~{\sc ii} & $\sim 2400$& $ 18 \pm 2$ & 37 & 20 & 38 &12 & 22\\
  Mg~{\sc ii} & 2798.74& $25 \pm 1$ & 50 & 24& 50 & 33$^b$& 64\\ \hline
\end{tabular}
}
\end{center}

\noindent $^a$\,\, Tentative identifications.

\noindent $^b$\,\, Profile incomplete.

\clearpage

\centerline{\bf FIGURE CAPTIONS}
\bigskip\parindent 0pt

Figure 1. -- Redshift distribution of sample quasars, with bin size of 0.33.
\bigskip

Figure 2. -- Lyman valley correction curves as a function of redshift
for lines with neutral--hydrogen column density between $2 \times 10^{12}$
and $10^{16}$~\cm.
\bigskip

Figure 3. -- Number of merging spectra as function of rest--frame wavelength.
Wavelength gaps between merging spectra and the exclusion of pixels with
very low S/N level cause  narrow features at short wavelengths.
\bigskip

Figure 4. -- Signal--to--noise level per \AA\ as function of rest--frame
wavelength. The actual level may be slightly lower due to uncertainties
in the FOS flat--field corrections. Correlated errors between adjacent
re--binned pixels are not taken into account.
\bigskip

Figure 5. --
Composite FOS spectrum of 101 quasars, binned to 2~\AA. Prominent emission
lines and the Lyman limit are labeled, and two possible emission features
are marked. The continuum fitting windows are marked with the bars near the
bottom.
\bigskip

Figure 6. --
Root--mean--square deviation per merging spectrum
normalized as a fraction of the composite spectrum in Fig.~5.
\bigskip

Figure 7. --
Spectral region near the Lyman limit. The spectrum is the same as Fig.~5, but
displayed with 0.3~\AA\ bins. The dotted line represents the
broken power-law continuum fitted in other windows (see \S 3.1).
The contributions to $\chi^2$ for each bin relative to the fitted continuum
are plotted in the lower panel.
\bigskip

Figure 8. -- Composite FOS spectra of 60 radio--loud and 41 radio--quiet
quasars, binned to 2~\AA. The flux level in the spectrum for radio--quiet
quasars is shifted down for display purposes.
Prominent emission lines and the Lyman limit are marked.
\bigskip

Figure 9. --
Comparison of quasar composite spectra. Solid curve: composite FOS spectrum of
radio--quiet quasars, representative of $\rm z \approx 1$;
dashed curve: optical composite spectrum of Francis \etal\ (1991),
representative of $\rm z \approx 3$.
The spectra are scaled so that the continuum levels around 1450~\AA\ in both
spectra match.
\bigskip

Figure 10. -- Model fit of Comptonized accretion disk to the composite FOS
spectrum. The composite spectrum is binned to 1~\AA\ and normalized to
the mean specific luminosity of the full sample.
The accretion disk model shown is a sum of blackbodies in the Schwarzschild
metric for a black hole mass of $1.4 \times 10^9\ M_\odot$, an accretion rate
of $2.8\ M_\odot\ yr^{-1}$, and a Lyman edge with optical depth 0.8.
The disk has an inclination of $30^\circ$. The assumed spherical Comptonizing
medium has an optical depth of 1.0 and a temperature of $4.1 \times 10^8$~K.
\bigskip

\clearpage

\begin{figure}
\plotone{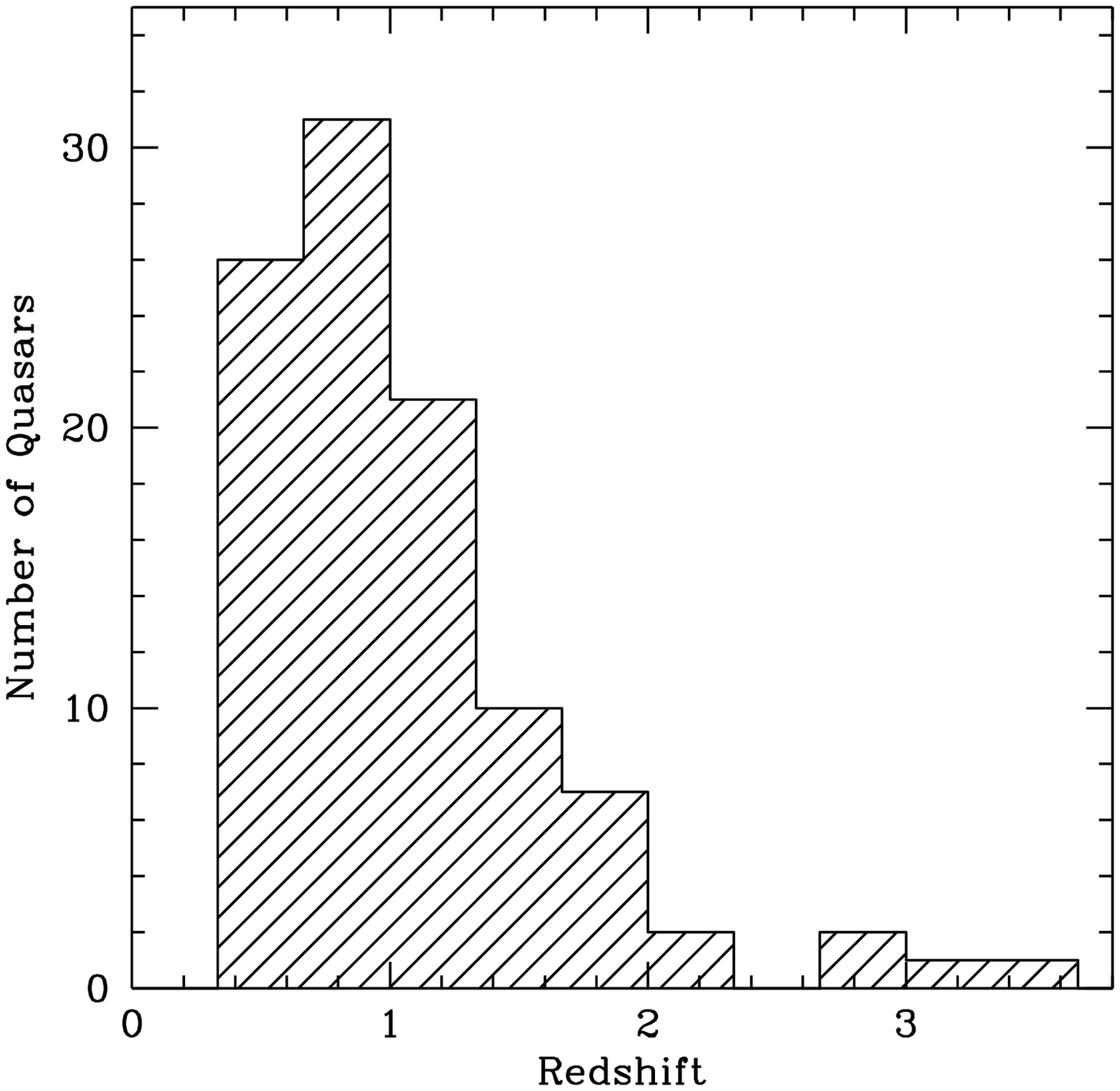}
\caption{~~}
\end{figure}

\begin{figure}
\plotone{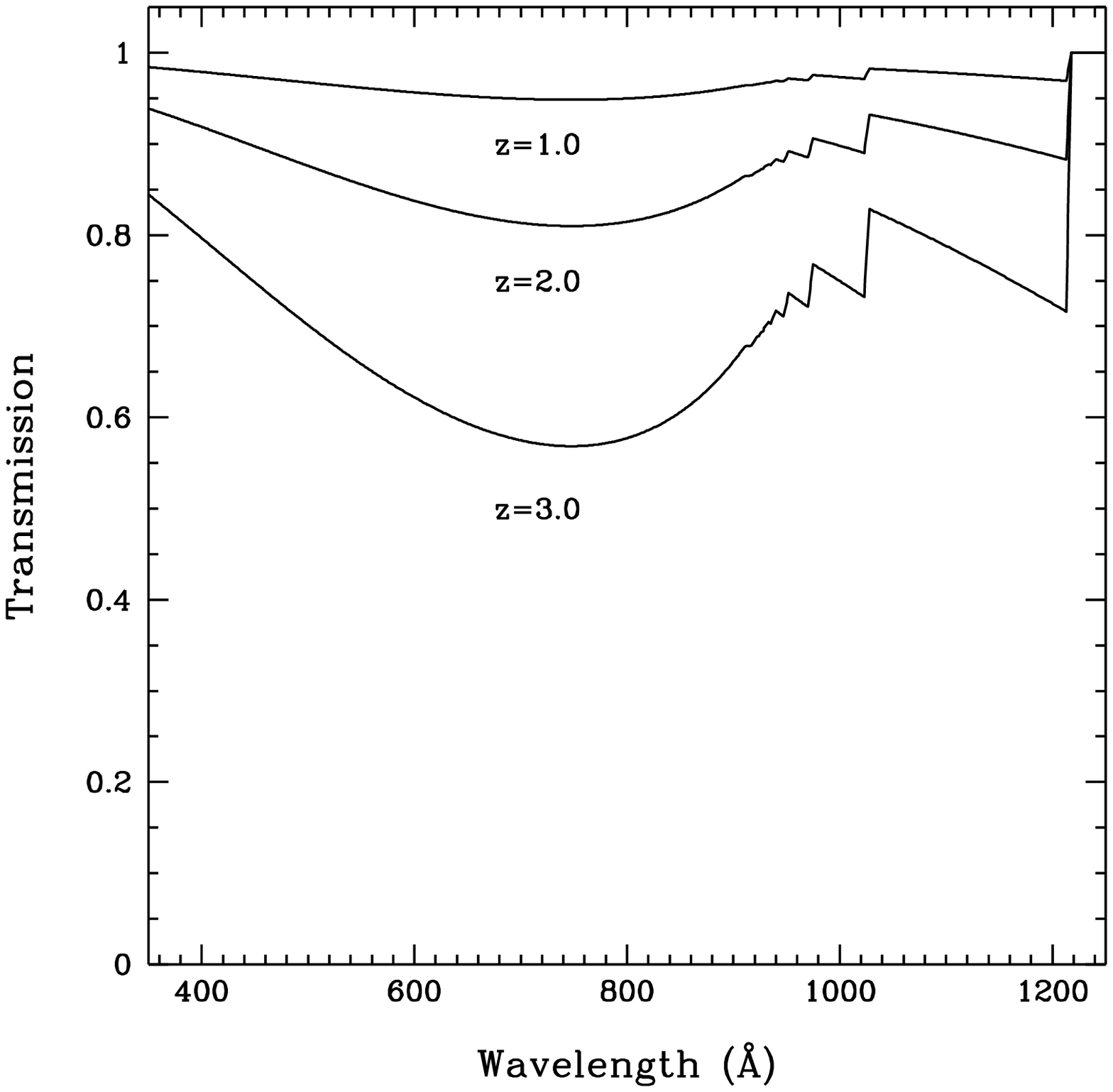}
\caption{~}
\end{figure}

\begin{figure}
\plotone{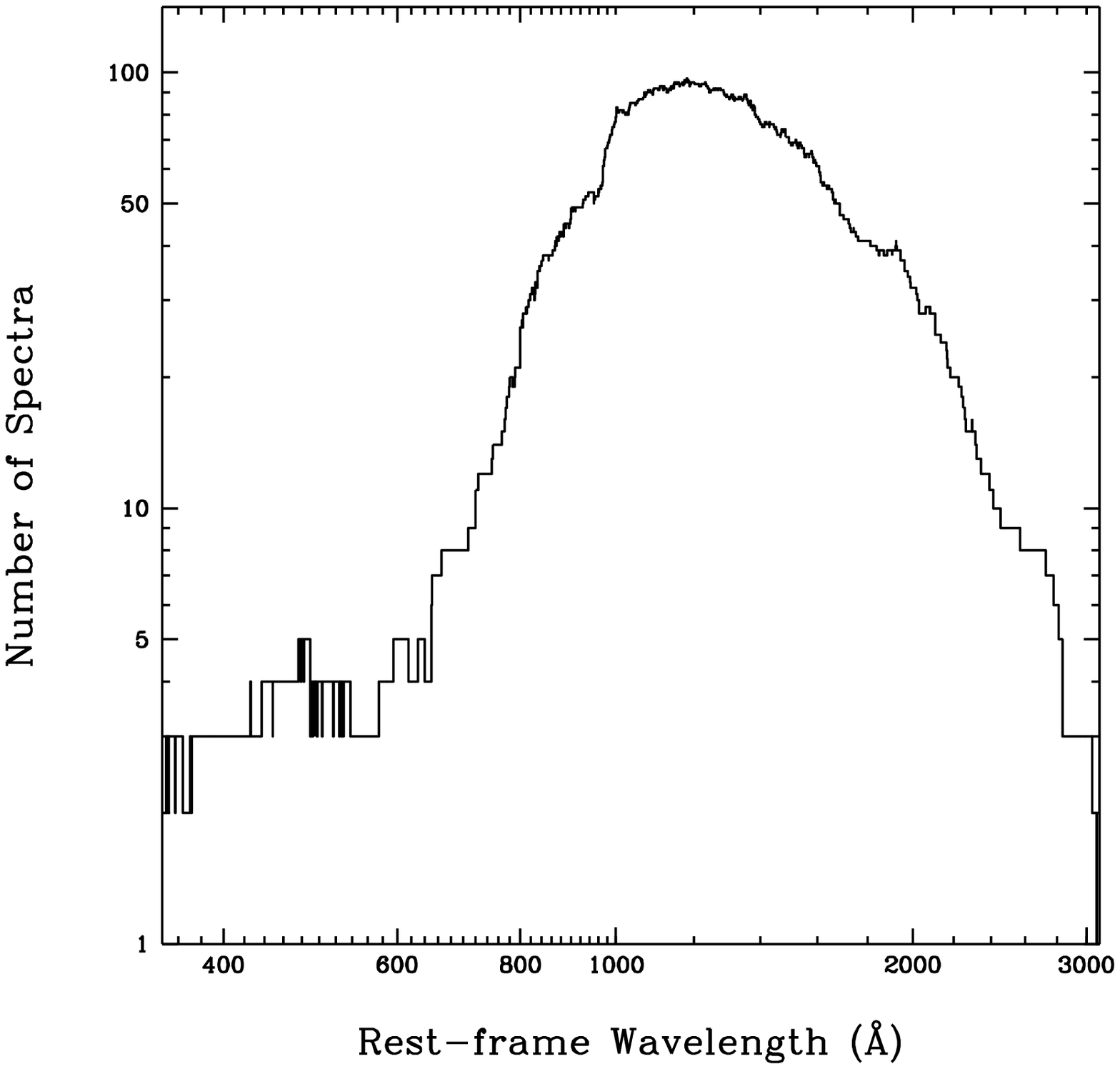}
\caption{~}
\end{figure}

\begin{figure}
\plotone{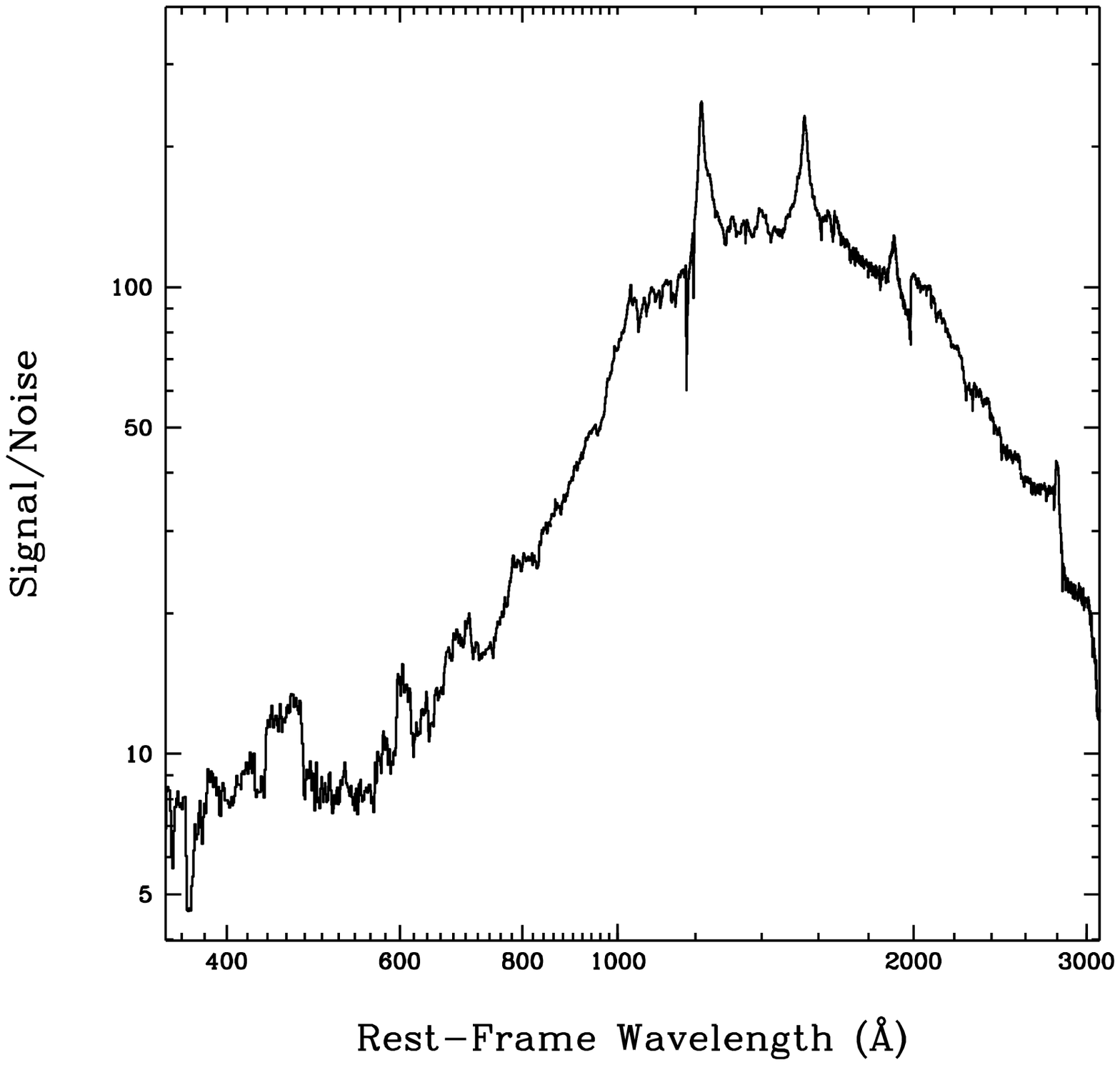}
\caption{~}
\end{figure}

\begin{figure}
\plotfiddle{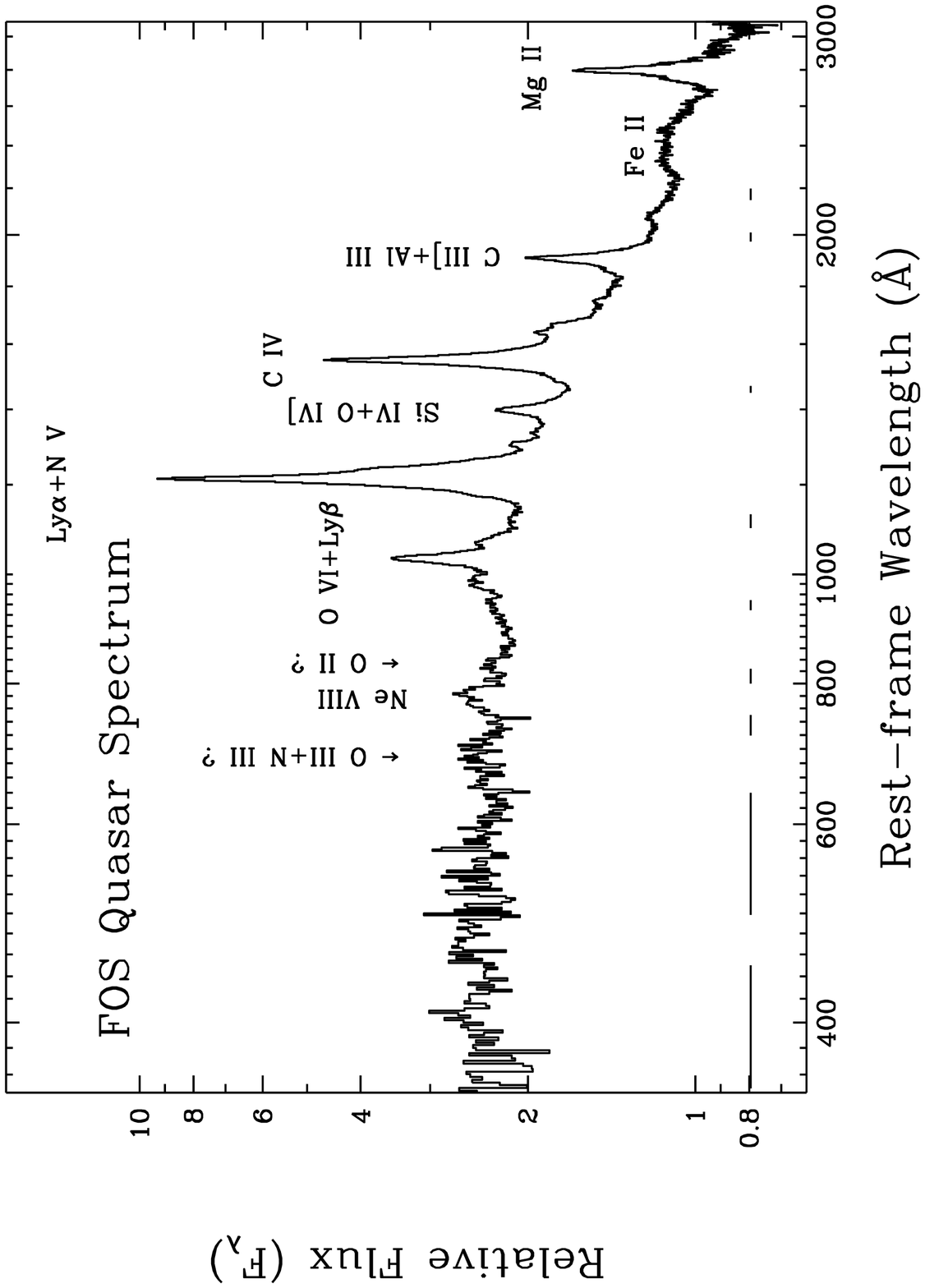}{7in}{-90}{75}{75}{-275}{500}
\caption{~}
\end{figure}

\begin{figure}
\plotone{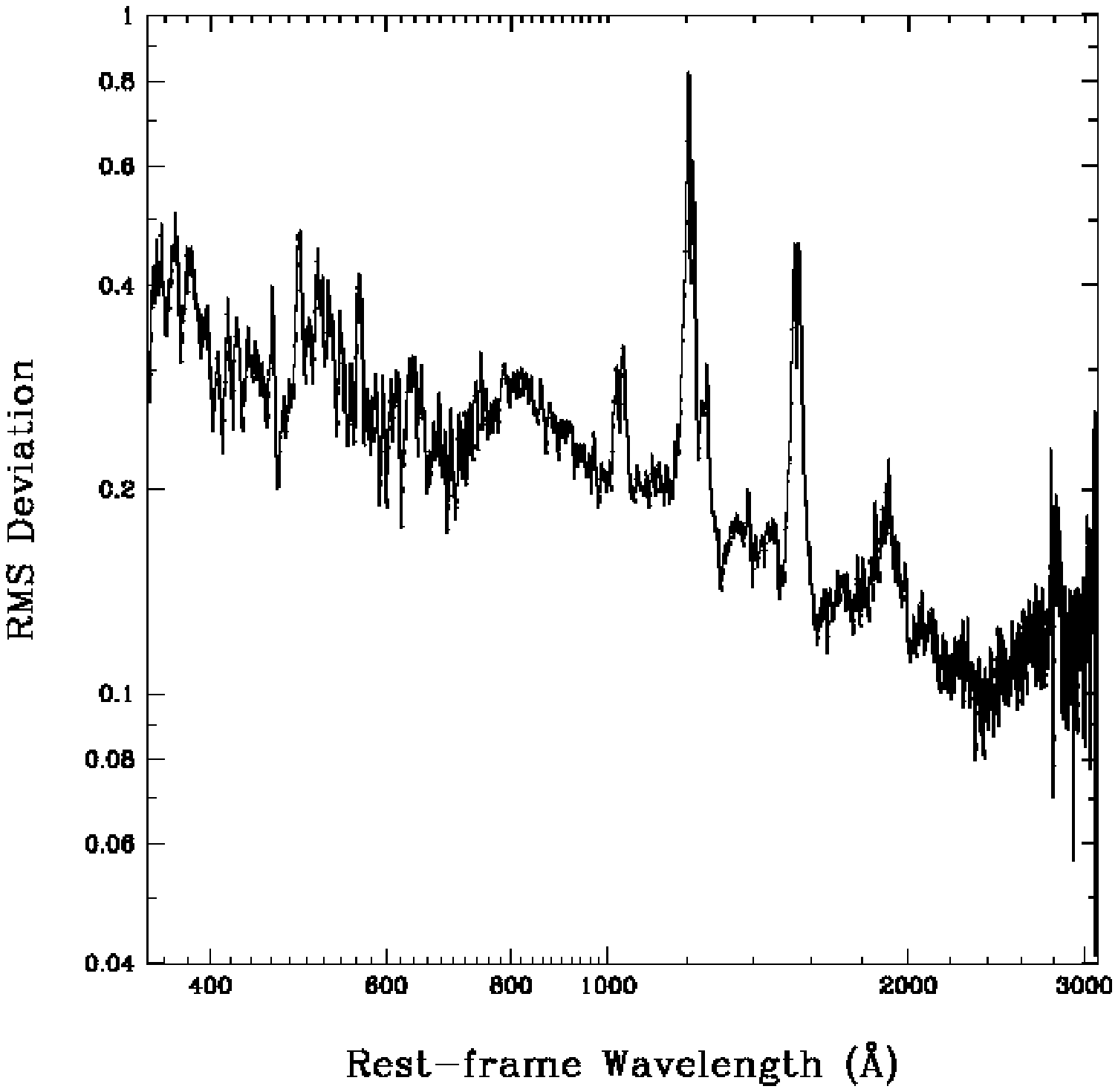}
\caption{~}
\end{figure}

\begin{figure}
\plotone{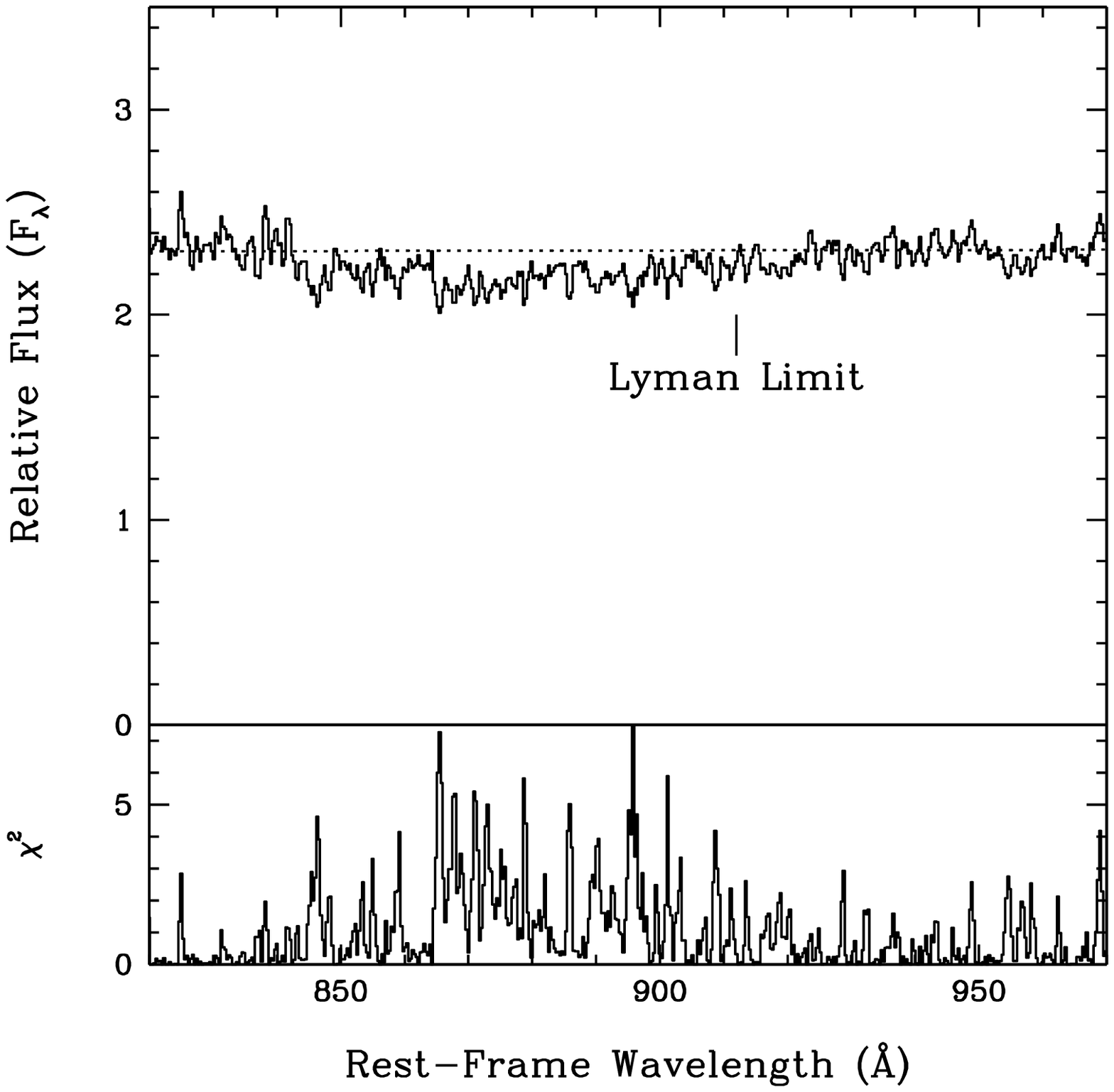}
\caption{~}
\end{figure}

\begin{figure}
\plotone{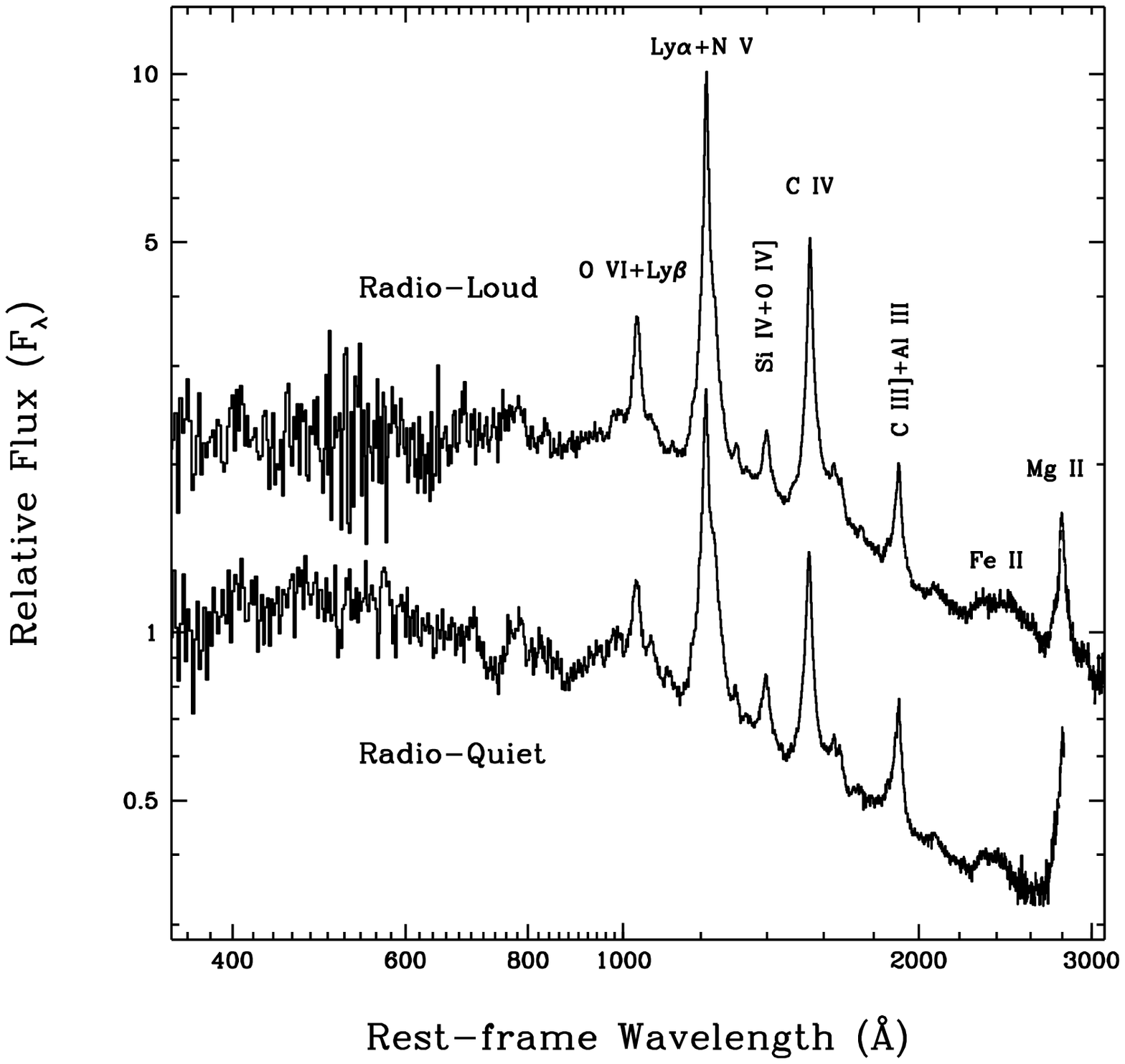}
\caption{~}
\end{figure}

\begin{figure}
\plotone{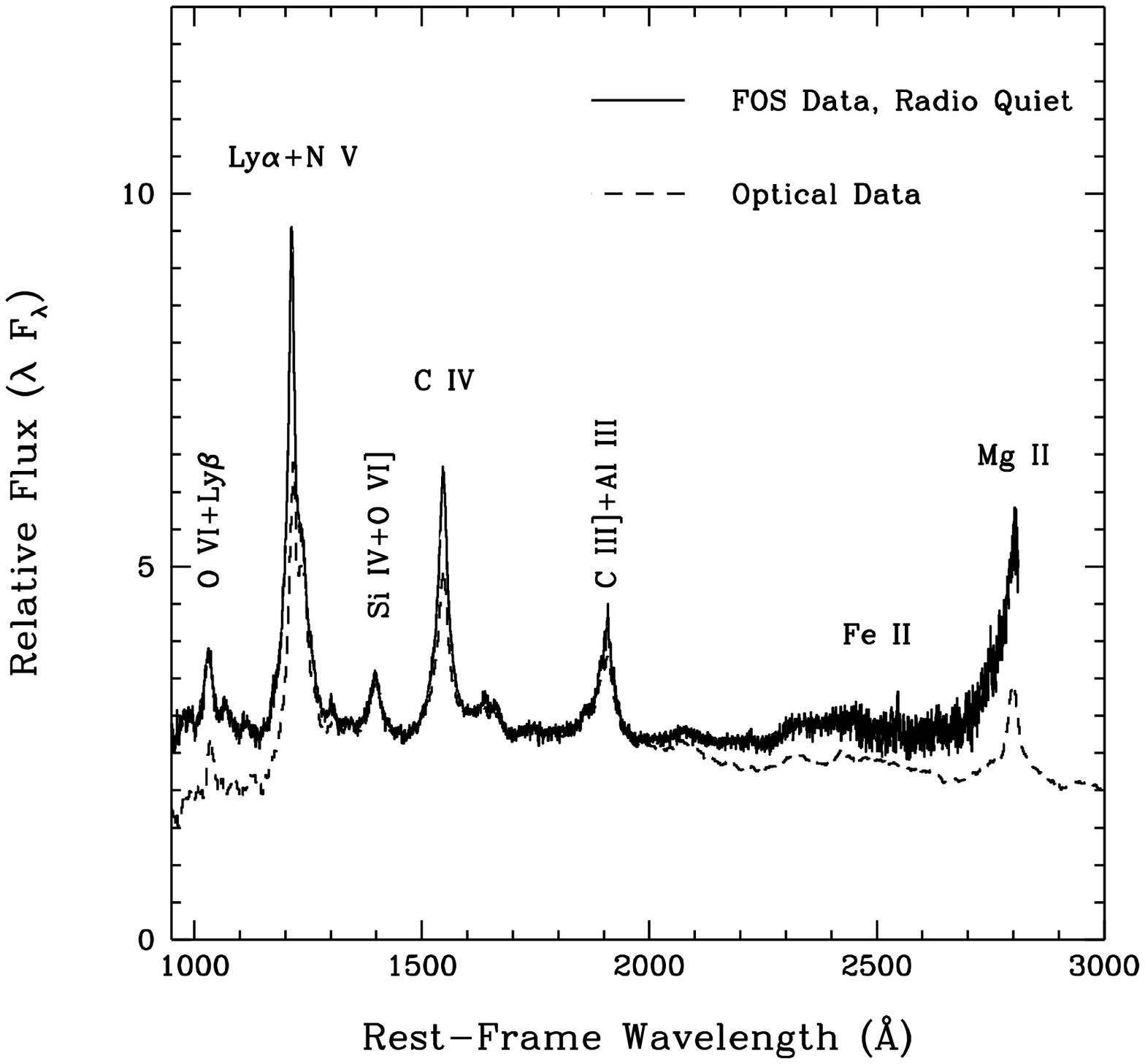}
\caption{~}
\end{figure}

\begin{figure}
\plotfiddle{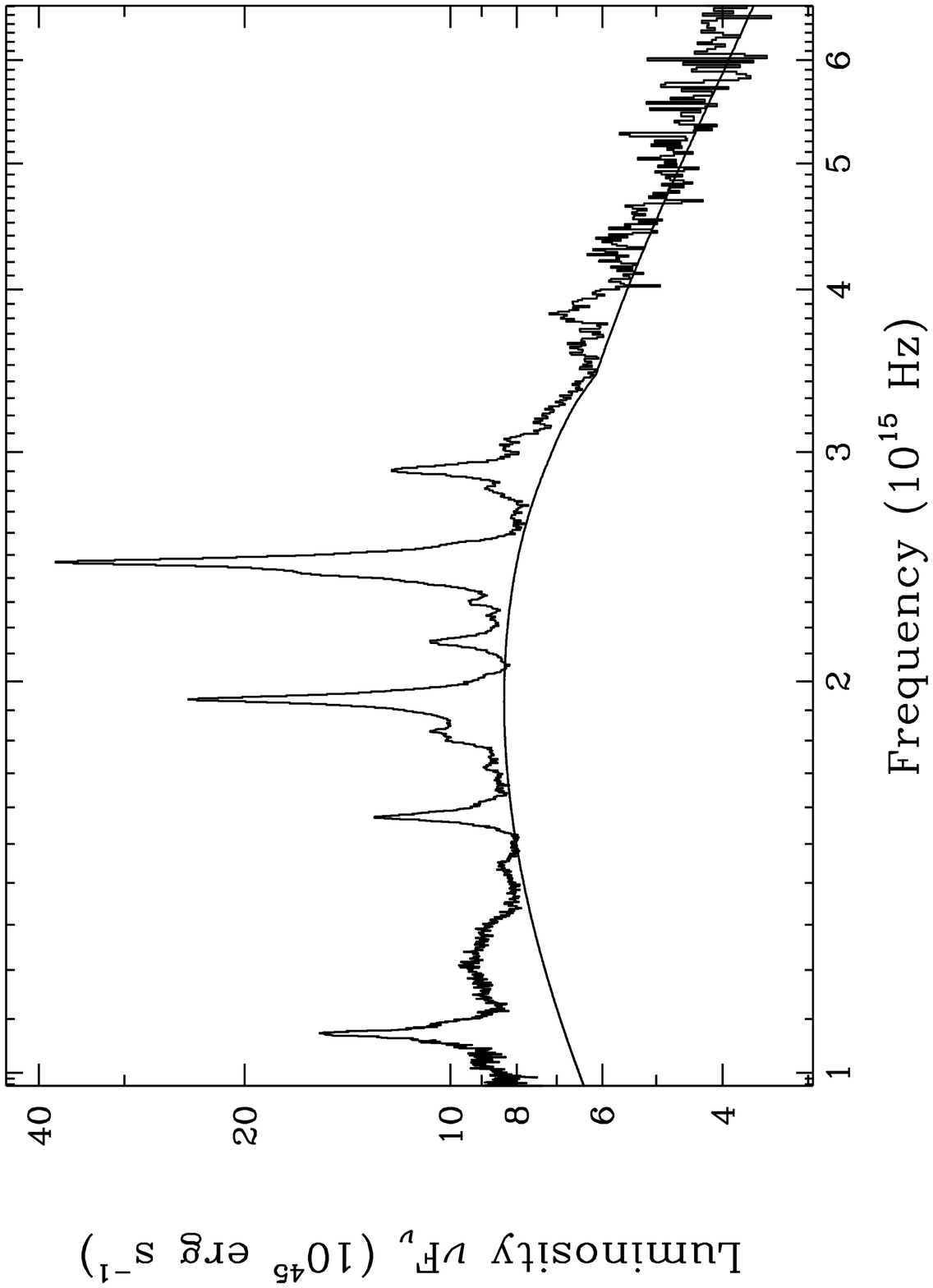}{7in}{-90}{75}{75}{-275}{500}
\caption{~}
\end{figure}
\end{document}